\documentclass[10pt, prl, preprintnumbers, aps, prl, twocolumn, floatfix]{revtex4-2}

\usepackage[english]{babel}
\usepackage{caption}
\usepackage{subcaption}
\usepackage[letterpaper, top=20mm, bottom=20mm, left=15mm, right=15mm, columnsep=15pt]{geometry}
\usepackage{amsmath}
\usepackage{amssymb}
\usepackage{graphicx}
\usepackage{graphics}
\usepackage[colorlinks=true, allcolors=blue]{hyperref}
\usepackage{physics}
\usepackage{bm}
\usepackage[dvipsnames]{xcolor}
\usepackage{ragged2e} 
\usepackage[normalem]{ulem} 
\usepackage{cancel} 

\newcommand{\slashed}[1]{#1\!\!\!/}

\begin{document}

\title{Fermion-selective tests of new physics with the bound-electron~$g$~factor}
\date{\today}
\author{M.~Moretti}
\affiliation{Max-Planck-Institut für Kernphysik, Heidelberg, Germany}
\author{C.~H.~Keitel}
\affiliation{Max-Planck-Institut für Kernphysik, Heidelberg, Germany}
\author{Z.~Harman}
\affiliation{Max-Planck-Institut für Kernphysik, Heidelberg, Germany}

\begin{abstract}
The use of high-precision measurements of the $g$ factor of single-electron ions is considered as a detailed probe for physics beyond the Standard Model. The contribution of the exchange of a hypothetical force-carrying scalar boson to the $g$ factor is calculated for the ground state of H-like ions and used to derive bounds on the parameters of that force. Similarly to the isotope shift, we employ the nuclide shift, i.e. the difference for elements with different proton and/or neutron numbers, in order to increase the experimental sensitivity to the new physics contribution. In particular we find, combining available measurements with current precision with different ions, that the coupling constant for the interaction between an electron and a proton can be constrained up to three orders of magnitude better than with the best current atomic data and theory.
\end{abstract}

\maketitle

The $g$ factors of the free electron and free muon have served as precision tests for quantum electrodynamics (QED), the Standard Model (SM) more broadly, and possible extensions of the SM~\cite{Fan2023,Volkov2019,CODATA2022}. Recent years have also seen rapid improvements in the experimental determination~\cite{Sturm2011,Sturm2013,Sturm2014,Heisse2023,Sailer2022} and theoretical prediction~\cite{Pachucki2004,Pachucki2017,Yerokhin2017,Karshenboim2005,Shabaev2002,Beier2000,Beier2000-1,Czarnecki2018,Glazov2019,Sikora2025} of the $g$ factor of bound electrons. This allows for stringent tests of QED in strong fields and, to a lesser extent, other sectors of the SM such as nuclear structural effects~\cite{Zatorski2012}, and can hence also be an avenue for the discovery of phenomena beyond the SM (BSM).

In the present Letter, we demonstrate the relevance of the $g$ factor of bound electrons in the search for physics beyond the SM.
Comparisons between the best available theoretical and experimental results, jointly with relative uncertainties, can provide competitive bounds for the NP contribution to the $g$ factor, induced, in this Letter, by a new massive scalar particle mediating a hypothetical force. With this method, we here constrain electron-neutron and electron-electron contributions independently. The same approach will also be used to constrain the electron-proton contribution with an orders-of-magnitude improvement, though not being fully independent of the other two.
Therefore, we also put forward an approach that generalizes the well-known isotope shift to isolate the contribution to the $g$ factor due to boson exchange between the bound electron and protons in the nucleus, the so-called \textit{nuclide shift}, i.e. the difference of $g$ factors for elements with different proton and/or neutron numbers. This quantity can be accessed with very high precision with modern experimental methods~\cite{Sailer2022}, as we discuss below.
We work in units of $c=1$ and $\hbar=1$.

\emph{Theory} --
We consider a scalar boson as a mediator of a hypothetical fifth force between electrons and nucleons.
The Lagrangian involving the scalar only reads~\cite{Srednicki2007}
\begin{equation}
    \mathcal{L}_\phi = -\frac{1}{2} \partial_\mu \phi \partial^\mu \phi - \frac{1}{2} m_\phi^2\phi^2 + y_e\phi\overline{\psi}_e\psi_e + \sum_N \, y_N\phi\overline{\psi}_N\psi_N  \,,
\end{equation}
where $\phi$ is the field associated to the scalar and the first two terms describe its kinematics, the subscripts $e$, $N \in \{n,p\}$ stand for electron and nucleon (neutron or proton), respectively. The $\psi$'s are the corresponding fermionic fields and the $y$'s are the coupling constants of this Yukawa theory. 
This induces three types of new possible interactions: electron-electron, electron-nucleon, and nucleon-nucleon. However, the latter is not considered in this work because it is included in the nuclear parameters extracted from experiments (\emph{e.g.} the nuclear radius and mass), and also our atomic quantities are not particularly sensitive to their value.

\begin{figure}
    \begin{subfigure}[b]{0.32\columnwidth}
        \includegraphics[width=\textwidth]{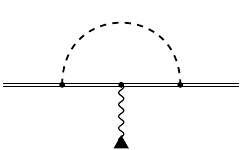}
        \caption{}
        \label{fig:g-factor SE bound irr}
    \end{subfigure}
    \hfill
    \begin{subfigure}[b]{0.32\columnwidth}
        \includegraphics[width=\textwidth]{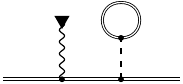}
        \caption{}
        \label{fig:g-factor VP bound red}
    \end{subfigure}
    \hfill
    \begin{subfigure}[b]{0.32\columnwidth}
        \includegraphics[width=\textwidth]{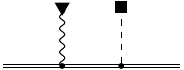}
        \caption{}
        \label{fig:g-factor eN}
    \end{subfigure}
    \vspace{0.1cm}
    \begin{subfigure}{0.32\columnwidth}
        \includegraphics[width=\linewidth]{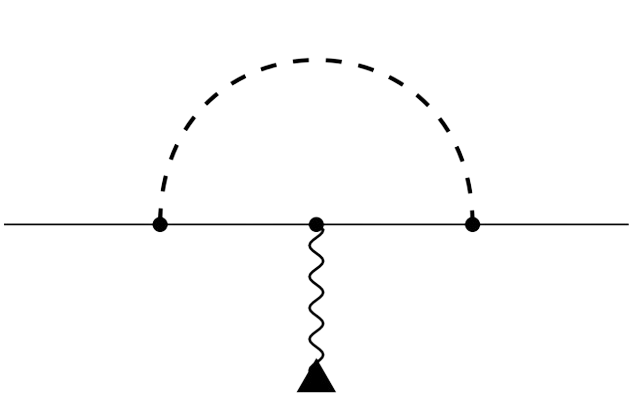}
        \caption{}
        \label{fig:g-factor SE free}
    \end{subfigure}
    \hfill
    \begin{subfigure}{0.32\columnwidth}
        \includegraphics[width=\linewidth]{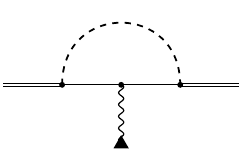}
        \caption{}
        \label{fig:g-factor SE first approx}
    \end{subfigure}
    \hfill
    \begin{subfigure}{0.32\columnwidth}
        \includegraphics[width=\linewidth]{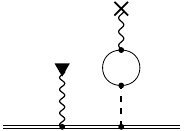}
        \caption{}
        \label{fig:g-factor VP first approx}
    \end{subfigure}
    \caption{\justifying
    First line: exemplary Feynman diagrams depicting perturbative scalar-boson corrections to the $g$ factor of the electron from self energy (\subref{fig:g-factor SE bound irr}), vacuum polarization (\subref{fig:g-factor VP bound red}) and electron-nucleon interaction (\subref{fig:g-factor eN}). 
    Second line: Bosonic vertex correction to the $g$ factor due to free-electron SE (\subref{fig:g-factor SE free}), the diagram contributing to the first non-vanishing $Z\alpha$ order in bound electron SE (\subref{fig:g-factor SE first approx}) and VP (\subref{fig:g-factor VP first approx}).
    A double line represents a Coulomb-Dirac wave function or propagator, a wavy line terminated by a triangle denotes a photon from the external magnetic field, the dashed line terminated by a square denotes a scalar from the nucleons, while the dashed line alone represents the scalar propagator, a wavy line terminated with a cross denotes a Coulomb interaction with the nucleus.}
    \label{fig:g-factor all together}
\end{figure}

In what follows, we consider an H-like ion, where only one 1$s$-electron orbits around the nucleus. The nucleons will be treated separately as neutrons or protons, as both are present in the systems considered.
We will label by $ee$, $en$ and $ep$ the interactions of the electron with itself, the neutrons and the protons, respectively. These three interactions contribute, among other quantities, to the $g$ factor, which can be expressed as
\begin{equation}
\label{eq:g-factor contribution sum}
    \Delta g_\phi = \Delta g_\phi^{ee} + \Delta g_\phi^{en} + \Delta g_\phi^{ep}\,.
\end{equation}
The first term of Eq.~\eqref{eq:g-factor contribution sum} encapsulates electron self-energy (SE) and vacuum polarization (VP), for which exemplary diagrams are shown in Fig.~\ref{fig:g-factor SE bound irr} and~\ref{fig:g-factor VP bound red}, while the second and the third terms correspond to Fig.~\ref{fig:g-factor eN}. Any contribution of this new scalar must satisfy
\begin{equation}
\label{eq:g-factor exp-th difference}
    \Delta g^i_\phi \leq g^\text{exp} - g^\text{th} \, ,
\end{equation}
where $\Delta g^i_\phi$ may represent any of the terms on the right-hand-side of Eq.~\eqref{eq:g-factor contribution sum}, $g^\text{exp} = g^\text{SM} + g^\text{BSM}$ and $g^\text{th} = g^\text{SM}$ are the experimentally measured and theoretical computed $g$ factors, the former including all the SM and BSM contributions and the latter the SM ones only.

Let us constrain these three contributions independently. In order to do so we anticipate the following: (i) the electron-electron coupling can be effectively constrained via the free-electron $g$ factor, and this can be transferred to the bound-electron contribution $\Delta g_\phi^{ee}$; (ii)  $\Delta g_\phi^{en}$ can be limited through isotope shifts (see \cite{Debierre2020}); and (iii) for $\Delta g_\phi^{ep}$ we put forward a method that cancels the $en$ contribution and exploits the accuracy found in the $ee$ part.

The electron-electron interaction manifests itself in the SE and VP corrections. In the first case, the bound electron interacts with itself, while in the latter case it exchanges a boson with a virtual electron-positron pair. As we will see, it is sufficient to consider the corresponding free-electron diagram. It is in fact convenient to treat the binding effect on the electron as a perturbation to the free-electron case, as the accuracy of the latter is better than the one of the former, and the binding correction is sufficiently small in the region of atomic numbers we are interested in.
The corrections we calculated are up to order of $(Z\alpha)^2$, and they stem from one-loop vertex diagram.

The free SE $g$ factor is derived from the $1$-loop correction of the QED vertex, with the virtual photon replaced by a scalar. Using the slash notation $\slashed{a} := a_\mu \gamma^\mu$, the Lagrangian reads~\cite{Srednicki2007}
\begin{multline}
    \mathcal{L} = \overline{\psi_e} \left( i\slashed{\partial} - e\slashed{A} - m_e \right)\psi_e -\frac{1}{4}F_{\mu\nu}F^{\mu\nu} + \\
    - \frac{1}{2} \partial_\mu\phi\partial^\mu\phi - \frac{1}{2}m_\phi^2\phi^2 + y_e\phi\overline{\psi}_e\psi_e \, ,
\end{multline}
where $A_\mu$ is the electromagnetic field and $F_{\mu\nu}$ is the corresponding strength field tensor. As the vertex $i e\gamma^\mu$ is corrected by QED (see~\cite{Srednicki2007}), it receives a correction from the scalar, too, as Fig.~\ref{fig:g-factor SE free} shows, transforming into the general vertex function
\begin{equation}
    i e \, \Gamma^\mu(p',p) = i e \left[ F_1(q^2) \gamma^\mu - \frac{i}{2m}F_2(q^2) \sigma^{\mu\nu} q_\nu \right] \, ,
\end{equation}
with $\sigma^{\mu\nu} = \frac{i}{2}\left[ \gamma^\mu, \gamma^\nu \right]$, when sandwiched between two free Dirac states of momenta $p'$ and $p$, and contracted with a photon of polarization $\epsilon^\mu(q)$. Here, $F_1(q^2)$ and $F_2(q^2)$ are the electron form factors and they are given by the sum of the contributions coming from the different propagators used to close the loops.
While $F_1$ modifies the original electric charge, $F_2$ leads to a modification of the magnetic moment, thus $g = 2 \left( 1 + F_2(0) \right)$. The one-scalar loop correction, analogous to QED calculations~\cite{Srednicki2007}, reads
\begin{multline}
    F_2^\phi (q^2) = \frac{y_e^2}{8\pi^2} \int_0^1 dx_1 dx_2 dx_3 \delta (1 - x_1 - x_2 - x_3) \times \\
    \times \frac{1-x_3^2}{(1-x_3)^2 + x_1 x_2 \frac{q^2}{m_e^2} + x_3 \frac{m_\phi^2}{m_e^2}} \, ,
\end{multline}
and for an on-shell external photon ($q^2 = 0$), the integral can be expressed in a closed form as
\begin{align}
    \nonumber
    F_2^\phi (0) & = \frac{\alpha_{ee}}{2\pi} \Bigg\{ \frac{3}{2} - r_m^2 -\frac{r_m \left( r_m^4 - 5r_m^2 +4 \right)}{\sqrt{r_m^2 - 4}} \, \times \\
    \nonumber
    & \times \left[ \text{atanh} \left( \frac{r_m^2 - 2}{r_m\sqrt{r_m^2 - 4}} \right) - \text{atanh} \left( \frac{r_m}{\sqrt{r_m^2 - 4}} \right) \right] \\
    \label{eq:F2(0)}
    & + r_m^2 \left( r_m^2 - 3 \right) \text{ln}\left( r_m \right) \Bigg\} \, ,
\end{align}
where $r_m = m_\phi/m_e$ is the ratio of the scalar and electron mass, and $\alpha_{ee} = y_e^2/4\pi$. The correction to the $g$ factor is then $\Delta g_\phi^{ee,\text{free}} = 2F_2^\phi (0)$.

For a H-like ion the only contributing diagram at leading order $(Z\alpha)^2$ is the one of Fig.~\ref{fig:g-factor SE first approx}.
As explained in e.g. Refs~.\cite{Hegstrom1973,Cakir2020}, this corresponds to computing the expectation value of the electron anomalous magnetic moment Hamiltonian $H = a_e \mu_B \gamma^0 \bm{B\cdot\Sigma}$ in a 1$s$ external state, namely:
\begin{equation}
\label{eq:Zeeman SE first approx}
    \bra{\psi_{1s}}H\ket{\psi_{1s}} = \mu_B B_z \, a_{e}^\phi \left[ 1-\frac{\left(Z\alpha\right)^2}{6} + O\left(\left(Z\alpha\right)^4\right) \right] \, ,
\end{equation}
with $\mu_B$ being the Bohr magneton, $B_z$ the projection of the magnetic potential onto the $z$-axis, and $a_{e}^\phi = F_2^\phi(0)$ the anomalous magnetic moment induced by the scalar. The contribution to the $g$ factor is
\begin{equation}
    \label{eq:g-factor first approx}
    \Delta g_\phi^{ee} = \Delta g_\phi^{ee,\text{free}} \left( 1 - \frac{(Z\alpha)^2}{6} \right) \, ,
\end{equation}
plus corrections of order $(Z\alpha)^4$ and higher orders.
To estimate the magnitude of $(Z\alpha)^4$ terms we consider one-loop SE and VP QED diagrams from Ref.~\cite{Pachucki2005}, with care taken to account for the different boson propagator. At one-loop level, the only photon propagator is replaced by a scalar propagator, such that our estimation amounts to the correspondent perturbation term of Eq. (51) of Ref.~\cite{Pachucki2005}, with the substitution $\frac{\alpha}{\pi} \rightarrow \frac{\alpha_{NP}}{\pi}$. It reads
\begin{align}
    \notag
    \Delta g_{\phi,(Z\alpha)^4}^{ee} \simeq \Delta g_\phi^{ee,\text{free}} & (Z\alpha)^4 \bigg[\frac{32}{9}\text{ln}\left( (Z\alpha)^{-2}\right) \\
    \label{eq:g-factor second approx}
    & - \frac{247}{216}-\frac{9}{8}\text{ln}(k_0)-\frac{8}{3}\text{ln}(k_3) \bigg] \, ,
\end{align}
omitting even higher orders in $Z\alpha$, where, for $1s$ states $\text{ln}(k_0) = 2.984\,128\,556$, and $\text{ln}(k_3) = 3.272\,806\,545$. Contributions up to $(Z\alpha)^4$ in Eq. \eqref{eq:g-factor second approx} are only small corrections to the free case for low and mid-Z nuclei ($<5\%$ for $Z = 50$, and $\sim 0.1 \%$ for $Z = 10$), which is sufficient for our purposes.

The best electron--neutron new physics exclusion limits in atomic precision spectroscopy originate from isotope shifts~\cite{Door2025,Ono2022,Sailer2022,Solaro2020}. The correction to the $g$ factor due to the new scalar mediator stems from the evaluation of the diagram in Fig.~\ref{fig:g-factor eN}, and it can be calculated by taking the following derivative~\cite{Karshenboim2005} of the expectation value of the Yukawa potential $V_\phi(\bm{r}) = -\alpha_{en} (A-Z) \frac{e^{-m_\phi |\bm{r}|}}{|\bm{r}|}$ experienced by the electron:
\begin{align}
    \Delta g_\phi^{en} = & \frac{\partial}{\partial m_e} \bra{\psi_{1s}} V_\phi\ket{\psi_{1s}} \nonumber \\
    \nonumber
    = & -\frac{4}{3}\alpha_{en} (A-Z) \frac{Z\alpha}{\gamma} \left( 1+ \frac{m_\phi}{2Z\alpha \, m_e} \right)^{-2\gamma} \times \\
    \label{eq:g-factor en}
    & \, \times \left[ 3 - 2\frac{\left( Z\alpha \right)^2}{1+\gamma} - 2\gamma \left( 1+ \frac{m_\phi}{2Z\alpha \, m_e} \right)^{-1}  \right] \, ,
\end{align}
where $\gamma=\sqrt{1-(Z\alpha)^2}$, $\alpha_{en} = y_e y_n/4\pi$ is the electron-neutron coupling and $A-Z$ in the number of neutrons in the nucleus. The above equation reproduces the results of~\cite{Debierre2020}, where the correspondent exclusion plot for $y_ey_n$ can be found. The contribution of $\Delta g_{\phi}^{ee}$ fully cancels in an isotope shift in our approximation~[see Eq.~\eqref{eq:g-factor first approx}].

As stated in Eq.~\eqref{eq:g-factor contribution sum}, the total $g$-factor modification is given by the sum of three different contributions, and because of Eq.~\eqref{eq:g-factor exp-th difference}, one could naively think that, in order to obtain the electron-proton part, it may be sufficient to isolate the $ep$ term in Eq.~\eqref{eq:g-factor contribution sum}, however, given that each piece of the right hand side of the latter equation undergoes a \emph{lower-than} inequality, a difference among them is not mathematically allowed to obtain an upper bound for $\Delta g_\phi^{ep}$. Either one relies on new methods to isolate $\Delta g_\phi^{ep}$ (that is what we propose later), or has to allow the least stringent scenario to take place, that is, $\Delta g_\phi^{ee} = \Delta g_\phi^{en} = 0$.
We follow this latter idea anyway because it nonetheless imposes a strong constraint on $y_e y_p$. The potential between the electron and nuclear protons has the very same form as the previous neutron one, and so does the $g$ factor correction for which the only differences compared to Eq.~\eqref{eq:g-factor en} are the replacements $\alpha_{en} \rightarrow \alpha_{ep} = y_e y_p /4\pi$ and $A-Z \rightarrow Z$, resulting in
\begin{multline}
    \Delta g_\phi^{ep} = -\frac{4}{3}\alpha_{ep} \, Z \, \frac{Z\alpha}{\gamma} \left( 1+ \frac{m_\phi}{2Z\alpha \, m_e} \right)^{-2\gamma} \times \\
    \label{eq:g-factor ep}
    \times \left[ 3 - 2\frac{\left( Z\alpha \right)^2}{1+\gamma} - 2\gamma \left( 1+ \frac{m_\phi}{2Z\alpha \, m_e}  \right)^{-1}  \right] \, .
\end{multline}
Along the lines of the isotope shift, where a cancellation of the electron-electron and electron-proton interactions takes place, here we put forward a method to cancel the neutron interaction and suppress the electronic self-interaction.
From Eq.~\eqref{eq:g-factor first approx} we see that the $ee$ contribution is proportional to the coupling constant via Eq.~\eqref{eq:F2(0)}, and the part inside the round brackets depends on the atomic number $Z$.
In Eqs. \eqref{eq:g-factor en} and \eqref{eq:g-factor ep}, we see a proportionality to the corresponding coupling constant, too, and since the two expressions have the same form except for the numbers $A-Z$ and $Z$, we rename them to $\Delta g_\phi^{en} = \alpha_{en}(A-Z)f(Z)$ and $\Delta g_\phi^{ep} = \alpha_{ep} \, Z \,f(Z)$, where $f(Z)$ is a function of $Z$ including every other remaining terms. Now, given two different ions, labeled $1$ and $2$, the $g$ factor corrections will vary depending on the number of protons and neutrons, namely:
\begin{align}
\nonumber
    \Delta g_{\phi,1} = \Delta g_\phi^{ee,\text{free}} & \left( 1 - \frac{(Z_1\alpha)^2}{6} \right) + \\
    \label{eq:g-factor ion1}
    & + \alpha_{en}(A_1-Z_1)f(Z_1) + \alpha_{ep} \, Z_1 \,f(Z_1) \, ,\\
\nonumber
    \Delta g_{\phi,2} = \Delta g_\phi^{ee,\text{free}} & \left( 1 - \frac{(Z_2\alpha)^2}{6} \right) +  \\
    \label{eq:g-factor ion2}
    & + \alpha_{en}(A_2-Z_2)f(Z_2) + \alpha_{ep} \, Z_2 \,f(Z_2) \, .
\end{align}
Subtracting from the former expression the latter multiplied by the quantity $x := \frac{A_1-Z_1}{A_2-Z_2} \frac{f(Z_1)}{f(Z_2)}$, dependent on $m_\phi$, the neutron part is completely canceled, and the electron-electron part is suppressed, resulting in
\begin{align}
\nonumber
    \Delta g_{\phi,1-2}^{\text{NS}} & = \Delta g_{\phi,1} - x \Delta g_{\phi,2} \\
\nonumber
    & = \Delta g_\phi^{ee,\text{free}} \left[ \left( 1 - \frac{(Z_1\alpha)^2}{6} \right) - x \left( 1 - \frac{(Z_2\alpha)^2}{6} \right) \right] \\
\label{eq:g-factor w diff}
    & \quad \, + y \, \alpha_{ep} f(Z_1)\,,
\end{align}
where we defined $y = Z_1 - Z_2\frac{A_1-Z_1}{A_2-Z_2} $. The upper index NS stands for ``nuclide shift".
Because an experiment can measure the $g$ factor of two ions, or a difference between them (as in the case of isotope shifts~\cite{Sailer2022}), we recast Eq.~\eqref{eq:g-factor w diff} into
\begin{align}
\nonumber
    \Delta g_{\phi,1-2}^{\text{NS}} & = \Delta g_{\phi,1} - \Delta g_{\phi,2} + (1-x)\Delta g_{\phi,2} \\
    \label{eq:g-factor w diff exp}
    & = \Delta g_{\phi,1-2} + (1-x)\Delta g_{\phi,2}\,,
\end{align}
where $\Delta g_{\phi,1-2} = \Delta g_{\phi,1} - \Delta g_{\phi,2}$ is the regular difference between the two $g$ factor contributions.
A dedicated experiment needs to measure both the $g$ factor difference and the value for a single ion, and, as recently, isotope shift $g$ factor experiments were shown to be more precise~\cite{Sailer2022} than those involving the $g$ factor of a single ion, one can anticipate a similar behavior for an experiment based on this nuclide shift method. To this end, a value $x \simeq 1$ (such that $1-x \ll 1$) is preferred to suppress inaccuracies due to $\Delta g_{\phi,2}$ in Eq.~(\ref{eq:g-factor w diff exp}). The question arises which pair of isotopes works best (more details can be found in the Supplementary Material~\cite{Supplemental_Material}).
\begin{figure*}[]
    \includegraphics[width=0.9\textwidth]{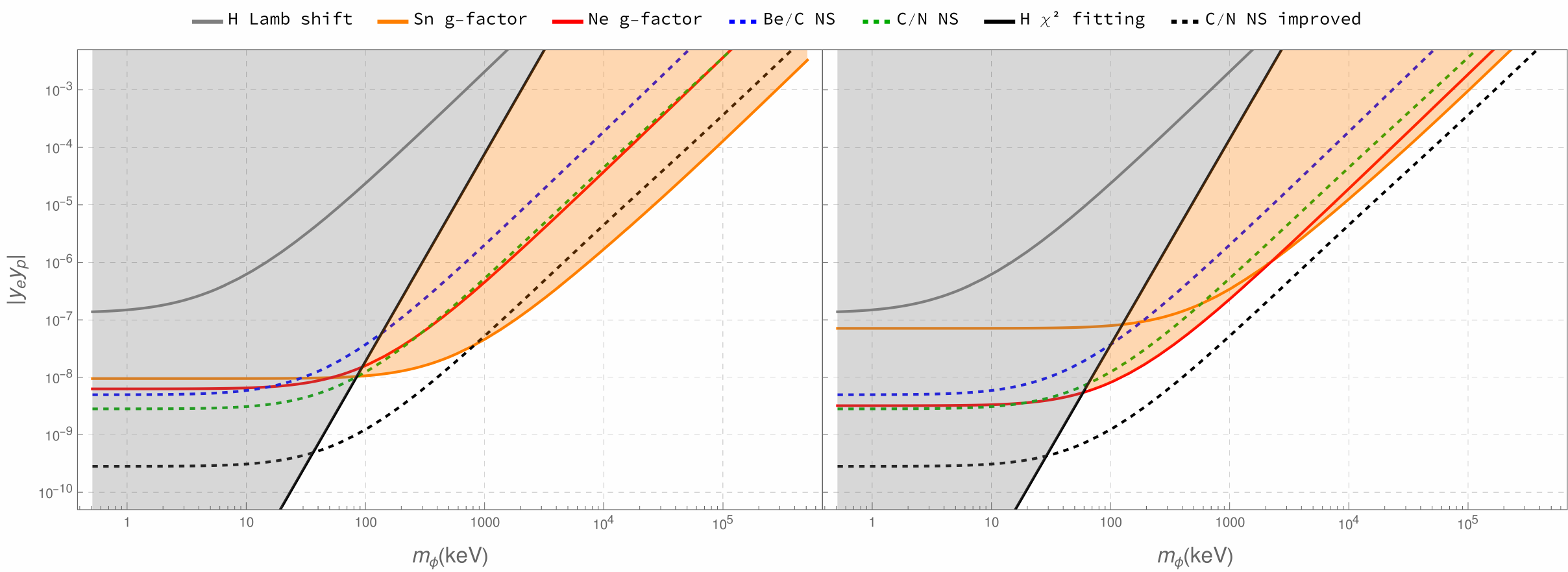}
    \caption{\justifying Exclusion lines for electron-proton attractive (left) and repulsive (right) coupling. Solid lines are based on experimental and theoretical available data, whereas dashed ones are projections.}
    \label{fig:yeypM}
\end{figure*}
In this work, we look for isotopes with a half-life of at least a few months and in the low-mid nuclear number range, such that $x \simeq 0.9$.
We find that the pair that optimally satisfies these requests is $^{14}\text{C}^{5+}$, $^{14}\text{N}^{6+}$ for which $x \in [0.72,0.98]$. Note that the electronic $g$ factor of an ion with nonzero nuclear spin can be expressed from the experimentally measurable $g_F$ factors of the hyperfine levels~\cite{Yerokhin2012} or from transition frequencies between magnetic sublevels~\cite{Moskovkin2006}.
Another suitable pair is $^{9}\text{Be}^{3+}$, $^{12}\text{C}^{5+}$ for which $x \in[0.24,0.56]$; this latter does not fulfill the requirement for $x$, but it is still very interesting for our purposes and sets competitive bounds (see the Supplementary Material~\cite{Supplemental_Material} for more details).
In order to give a proper bound on $\alpha_{ep}$ using this method, we need experimental and theoretical data. Since for some nuclides we have neither one nor the other, we derive projected bounds based on an estimation using data from the neon isotope shift experiment~\cite{Sailer2022, Heisse2023} from the beryllium theory~\cite{Sikora_private_comm2023}, and from carbon and nitrogen theory~\cite{Czarnecki2018}. The estimation of uncertainties is explained in the Supplementary material. 

\emph{Results} --
We found that the limit for $y_e y_e$ through the free-electron $g-2$ is already below $10^{-10}$ for small scalar masses $m_\phi$, and in our NS it becomes even more suppressed.\\
According to this model, exclusion plots for the coupling constant $y_e y_p$ are presented in Fig. \ref{fig:yeypM}, where the parameter regions above the curves are excluded.
For the sake of generality, we allow both the positive and negative signs of $y_e y_p$, translating into a repulsive and an attractive force, respectively.
To draw the exclusion curves, we compared experimental and theoretical data, and assumed that their difference within the error bars can be explained by the new physics contribution, as in Eq.~\eqref{eq:g-factor exp-th difference}. The resulting bounds are
obtained at one-sided 95\% confidence level.
The plots have been generated by making use of electronic wave functions corrected for a uniform spherical charge distribution of the nucleus, and the Yukawa potential has been convoluted with the same nuclear model. Nuclear RMS radius values have been taken from~\cite{Angeli2013}. Finite nuclear size corrections are practically negligible for our interests for light ions, \emph{e.g.} for neon ($Z=10$) they are $\sim1\%$ for $m_\phi\sim 1\,000m_e\sim 500\,000\,\text{keV}$, already decreasing to $\sim 0.001\%$ for $m_\phi\sim 100m_e\sim 50\,000\,\text{keV}$. However they grow significantly for tin ($Z=50$), which features a $\sim1\%$ correction for $m_\phi\sim 20m_e\sim 1\,000\,\text{keV}$, raising to $\sim42\%$ for $m_\phi\sim 1\,000m_e\sim 500\,000\,\text{keV}$.
\\Solid lines are based on fully available data and depend on the sign of $y_e y_p$, dashed lines are projections, independent on $y_e y_p$. The gray ones come from the hydrogen $1s-2s$ Lamb shift data~\cite{Matveev2013,Yerokhin2015}, where we employed theory from~\cite{Debierre2020}. The solid black curves are linear extrapolations of hydrogen data from Refs.~\cite{Potvliege2025,Jones2020}, where the exclusion plots have been drawn through $\chi^2$--fitting using many transitions in the hydrogen atom. It establishes a competitive bound in the mass range up to~30 keV, after which, hydrogen stops performing well. Under the stricter physical assumption that a muon couples to a proton with the same coupling constant as an electron, Ref.~\cite{Potvliege2025} reports an even better curve, although this case is not directly comparable to ours.
Orange and red curves are consequence of Eq. \eqref{eq:g-factor ep}.
The orange curves stem from the mid-$Z$ H-like tin $g$ factor measurement~\cite{Morgner2023}, for which the QED theory has recently been improved by a nonperturbative calculation of the two-loop self-energy corrections~\cite{Sikora2025}; although $g$ factor experimental uncertainties remain approximately the same as the charge number increases, the same cannot be said for theoretical uncertainties, since certain higher-order corrections become more and more important with increasing $Z$. This is reflected in the red curves, for the low-$Z$ neon ion, where a smaller theoretical error bar yields a better exclusion line~\cite{Heisse2023} in the mid-low mass regime. However, the higher nuclear number of Sn remains a better probe for the low-range/high-mass regime.
Next, we point to the blue and green dashed lines, for which the NS method has been employed. As mentioned above, to draw these lines we approximated the experimental error bar of the NS to be the same as that of the isotope shift in the experiment~\cite{Sailer2022}, and assumed theoretical error bars from \cite{Sikora_private_comm2023} for Be, and \cite{Czarnecki2018} for C and N.
Lacking values for some elements, we assume that QED theoretical and experimental values are equal.
The combination of dashed lines with solid ones generates, in each plot, the most competitive bounds for the mid- and high-mass regime, enhancing bounds by at least two orders of magnitude for boson masses above 1~MeV.
Finally, the black, dashed curves have been generated similarly to the green ones, with the assumption of a feasible improvement~\cite{Sturm_private_comm2025} of one order of magnitude in the experimental accuracy with respect to the isotope shift mentioned above~\cite{Sailer2022} and assuming a matching theoretical improvement.
The orange shaded area presents the improvement achieved with the available data compared to the best current bounds~\cite{Jones2020,Potvliege2025} shown by the gray shaded area. Further improvements could be obtained with the NS method, as indicated by the dashed lines.

\emph{Conclusions and outlook} --
\label{sec:discussion}
In this Letter we analyzed the contribution of a hypothetical scalar boson to the bound-electron $g$ factor in hydrogen-like ions. Beside the electron self-interaction and electron-nucleon interactions, we put forward a method to selectively probe the electron-proton coupling in a semi-independent way which is otherwise not accessible in other spectroscopic studies. This results in a series of competitive exclusion plots of coupling strength against the mass of the scalar.
An effective disentanglement of the three contributions has been performed by employing a $g$ factor shift between two ions with different nuclear charges, and identifying the optimal pair of nuclides. Projections of Fig.~\ref{fig:yeypM} follow from existing theoretical data and foreseeable near-future ``nuclide shift" experiments generalizing the concept of the well-known isotope shift, with the reasonable assumption of having experimental error bars similar to those already achieved.
With our concept, we were able to provide and predict exclusion curves for electron-proton interaction with stringency comparable to that of the electron-neutron interaction, improving the best previous bounds 
up to 3 orders of magnitude in the regime of high boson masses above 100~keV.
Near-future experimental advancements in Penning trap measurements, combined with foreseeable improved theoretical calculations of QED and nuclear structure effects, will further enhance the sensitivity of these tests.
\\

\emph{Acknowledgments} — We thank Konstantin A. Beyer for insightful discussions. Supported by the Deutsche Forschungsgemeinschaft (DFG, German Research Foundation) Project-ID 273811115—SFB 1225.

%


\begin{thebibliography}{38}%
\makeatletter
\providecommand \@ifxundefined [1]{%
 \@ifx{#1\undefined}
}%
\providecommand \@ifnum [1]{%
 \ifnum #1\expandafter \@firstoftwo
 \else \expandafter \@secondoftwo
 \fi
}%
\providecommand \@ifx [1]{%
 \ifx #1\expandafter \@firstoftwo
 \else \expandafter \@secondoftwo
 \fi
}%
\providecommand \natexlab [1]{#1}%
\providecommand \enquote  [1]{``#1''}%
\providecommand \bibnamefont  [1]{#1}%
\providecommand \bibfnamefont [1]{#1}%
\providecommand \citenamefont [1]{#1}%
\providecommand \href@noop [0]{\@secondoftwo}%
\providecommand \href [0]{\begingroup \@sanitize@url \@href}%
\providecommand \@href[1]{\@@startlink{#1}\@@href}%
\providecommand \@@href[1]{\endgroup#1\@@endlink}%
\providecommand \@sanitize@url [0]{\catcode `\\12\catcode `\$12\catcode
  `\&12\catcode `\#12\catcode `\^12\catcode `\_12\catcode `\%12\relax}%
\providecommand \@@startlink[1]{}%
\providecommand \@@endlink[0]{}%
\providecommand \url  [0]{\begingroup\@sanitize@url \@url }%
\providecommand \@url [1]{\endgroup\@href {#1}{\urlprefix }}%
\providecommand \urlprefix  [0]{URL }%
\providecommand \Eprint [0]{\href }%
\providecommand \doibase [0]{http://dx.doi.org/}%
\providecommand \selectlanguage [0]{\@gobble}%
\providecommand \bibinfo  [0]{\@secondoftwo}%
\providecommand \bibfield  [0]{\@secondoftwo}%
\providecommand \translation [1]{[#1]}%
\providecommand \BibitemOpen [0]{}%
\providecommand \bibitemStop [0]{}%
\providecommand \bibitemNoStop [0]{.\EOS\space}%
\providecommand \EOS [0]{\spacefactor3000\relax}%
\providecommand \BibitemShut  [1]{\csname bibitem#1\endcsname}%
\let\auto@bib@innerbib\@empty
\bibitem [{\citenamefont {Fan}\ \emph {et~al.}(2023)\citenamefont {Fan},
  \citenamefont {Myers}, \citenamefont {Sukra},\ and\ \citenamefont
  {Gabrielse}}]{Fan2023}%
  \BibitemOpen
  \bibfield  {author} {\bibinfo {author} {\bibfnamefont {X.}~\bibnamefont
  {Fan}}, \bibinfo {author} {\bibfnamefont {T.~G.}\ \bibnamefont {Myers}},
  \bibinfo {author} {\bibfnamefont {B.~A.~D.}\ \bibnamefont {Sukra}}, \ and\
  \bibinfo {author} {\bibfnamefont {G.}~\bibnamefont {Gabrielse}},\ }\href
  {\doibase 10.1103/PhysRevLett.130.071801} {\bibfield  {journal} {\bibinfo
  {journal} {Phys. Rev. Lett.}\ }\textbf {\bibinfo {volume} {130}},\ \bibinfo
  {pages} {071801} (\bibinfo {year} {2023})}\BibitemShut {NoStop}%
\bibitem [{\citenamefont {Volkov}(2019)}]{Volkov2019}%
  \BibitemOpen
  \bibfield  {author} {\bibinfo {author} {\bibfnamefont {S.}~\bibnamefont
  {Volkov}},\ }\href {\doibase 10.1103/PhysRevD.100.096004} {\bibfield
  {journal} {\bibinfo  {journal} {Phys. Rev. D}\ }\textbf {\bibinfo {volume}
  {100}},\ \bibinfo {pages} {096004} (\bibinfo {year} {2019})}\BibitemShut
  {NoStop}%
\bibitem [{\citenamefont {Mohr}\ \emph {et~al.}(2025)\citenamefont {Mohr},
  \citenamefont {Newell}, \citenamefont {Taylor},\ and\ \citenamefont
  {Tiesinga}}]{CODATA2022}%
  \BibitemOpen
  \bibfield  {author} {\bibinfo {author} {\bibfnamefont {P.~J.}\ \bibnamefont
  {Mohr}}, \bibinfo {author} {\bibfnamefont {D.~B.}\ \bibnamefont {Newell}},
  \bibinfo {author} {\bibfnamefont {B.~N.}\ \bibnamefont {Taylor}}, \ and\
  \bibinfo {author} {\bibfnamefont {E.}~\bibnamefont {Tiesinga}},\ }\href
  {\doibase 10.1103/RevModPhys.97.025002} {\bibfield  {journal} {\bibinfo
  {journal} {Rev. Mod. Phys.}\ }\textbf {\bibinfo {volume} {97}},\ \bibinfo
  {pages} {025002} (\bibinfo {year} {2025})}\BibitemShut {NoStop}%
\bibitem [{\citenamefont {Sturm}\ \emph {et~al.}(2011)\citenamefont {Sturm},
  \citenamefont {Wagner}, \citenamefont {Schabinger}, \citenamefont {Zatorski},
  \citenamefont {Harman}, \citenamefont {Quint}, \citenamefont {Werth},
  \citenamefont {Keitel},\ and\ \citenamefont {Blaum}}]{Sturm2011}%
  \BibitemOpen
  \bibfield  {author} {\bibinfo {author} {\bibfnamefont {S.}~\bibnamefont
  {Sturm}}, \bibinfo {author} {\bibfnamefont {A.}~\bibnamefont {Wagner}},
  \bibinfo {author} {\bibfnamefont {B.}~\bibnamefont {Schabinger}}, \bibinfo
  {author} {\bibfnamefont {J.}~\bibnamefont {Zatorski}}, \bibinfo {author}
  {\bibfnamefont {Z.}~\bibnamefont {Harman}}, \bibinfo {author} {\bibfnamefont
  {W.}~\bibnamefont {Quint}}, \bibinfo {author} {\bibfnamefont
  {G.}~\bibnamefont {Werth}}, \bibinfo {author} {\bibfnamefont {C.~H.}\
  \bibnamefont {Keitel}}, \ and\ \bibinfo {author} {\bibfnamefont
  {K.}~\bibnamefont {Blaum}},\ }\href {\doibase 10.1103/PhysRevLett.107.023002}
  {\bibfield  {journal} {\bibinfo  {journal} {Phys. Rev. Lett.}\ }\textbf
  {\bibinfo {volume} {107}},\ \bibinfo {pages} {023002} (\bibinfo {year}
  {2011})}\BibitemShut {NoStop}%
\bibitem [{\citenamefont {Sturm}\ \emph {et~al.}(2013)\citenamefont {Sturm},
  \citenamefont {Wagner}, \citenamefont {Kretzschmar}, \citenamefont {Quint},
  \citenamefont {Werth},\ and\ \citenamefont {Blaum}}]{Sturm2013}%
  \BibitemOpen
  \bibfield  {author} {\bibinfo {author} {\bibfnamefont {S.}~\bibnamefont
  {Sturm}}, \bibinfo {author} {\bibfnamefont {A.}~\bibnamefont {Wagner}},
  \bibinfo {author} {\bibfnamefont {M.}~\bibnamefont {Kretzschmar}}, \bibinfo
  {author} {\bibfnamefont {W.}~\bibnamefont {Quint}}, \bibinfo {author}
  {\bibfnamefont {G.}~\bibnamefont {Werth}}, \ and\ \bibinfo {author}
  {\bibfnamefont {K.}~\bibnamefont {Blaum}},\ }\href {\doibase
  10.1103/PhysRevA.87.030501} {\bibfield  {journal} {\bibinfo  {journal} {Phys.
  Rev. A}\ }\textbf {\bibinfo {volume} {87}},\ \bibinfo {pages} {030501}
  (\bibinfo {year} {2013})}\BibitemShut {NoStop}%
\bibitem [{\citenamefont {Sturm}\ \emph {et~al.}(2014)\citenamefont {Sturm},
  \citenamefont {K{\"o}hler}, \citenamefont {Zatorski}, \citenamefont {Wagner},
  \citenamefont {Harman}, \citenamefont {Werth}, \citenamefont {Quint},
  \citenamefont {Keitel},\ and\ \citenamefont {Blaum}}]{Sturm2014}%
  \BibitemOpen
  \bibfield  {author} {\bibinfo {author} {\bibfnamefont {S.}~\bibnamefont
  {Sturm}}, \bibinfo {author} {\bibfnamefont {F.}~\bibnamefont {K{\"o}hler}},
  \bibinfo {author} {\bibfnamefont {J.}~\bibnamefont {Zatorski}}, \bibinfo
  {author} {\bibfnamefont {A.}~\bibnamefont {Wagner}}, \bibinfo {author}
  {\bibfnamefont {Z.}~\bibnamefont {Harman}}, \bibinfo {author} {\bibfnamefont
  {G.}~\bibnamefont {Werth}}, \bibinfo {author} {\bibfnamefont
  {W.}~\bibnamefont {Quint}}, \bibinfo {author} {\bibfnamefont {C.~H.}\
  \bibnamefont {Keitel}}, \ and\ \bibinfo {author} {\bibfnamefont
  {K.}~\bibnamefont {Blaum}},\ }\href {\doibase 10.1038/nature13026} {\bibfield
   {journal} {\bibinfo  {journal} {Nature}\ }\textbf {\bibinfo {volume}
  {506}},\ \bibinfo {pages} {467} (\bibinfo {year} {2014})}\BibitemShut
  {NoStop}%
\bibitem [{\citenamefont {Hei\ss{}e}\ \emph {et~al.}(2023)\citenamefont
  {Hei\ss{}e}, \citenamefont {Door}, \citenamefont {Sailer}, \citenamefont
  {Filianin}, \citenamefont {Herkenhoff}, \citenamefont {K\"onig},
  \citenamefont {Kromer}, \citenamefont {Lange}, \citenamefont {Morgner},
  \citenamefont {Rischka}, \citenamefont {Schweiger}, \citenamefont {Tu},
  \citenamefont {Novikov}, \citenamefont {Eliseev}, \citenamefont {Sturm},\
  and\ \citenamefont {Blaum}}]{Heisse2023}%
  \BibitemOpen
  \bibfield  {author} {\bibinfo {author} {\bibfnamefont {F.}~\bibnamefont
  {Hei\ss{}e}}, \bibinfo {author} {\bibfnamefont {M.}~\bibnamefont {Door}},
  \bibinfo {author} {\bibfnamefont {T.}~\bibnamefont {Sailer}}, \bibinfo
  {author} {\bibfnamefont {P.}~\bibnamefont {Filianin}}, \bibinfo {author}
  {\bibfnamefont {J.}~\bibnamefont {Herkenhoff}}, \bibinfo {author}
  {\bibfnamefont {C.~M.}\ \bibnamefont {K\"onig}}, \bibinfo {author}
  {\bibfnamefont {K.}~\bibnamefont {Kromer}}, \bibinfo {author} {\bibfnamefont
  {D.}~\bibnamefont {Lange}}, \bibinfo {author} {\bibfnamefont
  {J.}~\bibnamefont {Morgner}}, \bibinfo {author} {\bibfnamefont
  {A.}~\bibnamefont {Rischka}}, \bibinfo {author} {\bibfnamefont
  {C.}~\bibnamefont {Schweiger}}, \bibinfo {author} {\bibfnamefont
  {B.}~\bibnamefont {Tu}}, \bibinfo {author} {\bibfnamefont {Y.~N.}\
  \bibnamefont {Novikov}}, \bibinfo {author} {\bibfnamefont {S.}~\bibnamefont
  {Eliseev}}, \bibinfo {author} {\bibfnamefont {S.}~\bibnamefont {Sturm}}, \
  and\ \bibinfo {author} {\bibfnamefont {K.}~\bibnamefont {Blaum}},\ }\href
  {\doibase 10.1103/PhysRevLett.131.253002} {\bibfield  {journal} {\bibinfo
  {journal} {Phys. Rev. Lett.}\ }\textbf {\bibinfo {volume} {131}},\ \bibinfo
  {pages} {253002} (\bibinfo {year} {2023})}\BibitemShut {NoStop}%
\bibitem [{\citenamefont {Sailer}\ \emph {et~al.}(2022)\citenamefont {Sailer},
  \citenamefont {Debierre}, \citenamefont {Harman}, \citenamefont {Heiße},
  \citenamefont {König}, \citenamefont {Morgner}, \citenamefont {Tu},
  \citenamefont {Volotka}, \citenamefont {Keitel}, \citenamefont {Blaum},\ and\
  \citenamefont {Sturm}}]{Sailer2022}%
  \BibitemOpen
  \bibfield  {author} {\bibinfo {author} {\bibfnamefont {T.}~\bibnamefont
  {Sailer}}, \bibinfo {author} {\bibfnamefont {V.}~\bibnamefont {Debierre}},
  \bibinfo {author} {\bibfnamefont {Z.}~\bibnamefont {Harman}}, \bibinfo
  {author} {\bibfnamefont {F.}~\bibnamefont {Heiße}}, \bibinfo {author}
  {\bibfnamefont {C.}~\bibnamefont {König}}, \bibinfo {author} {\bibfnamefont
  {J.}~\bibnamefont {Morgner}}, \bibinfo {author} {\bibfnamefont
  {B.}~\bibnamefont {Tu}}, \bibinfo {author} {\bibfnamefont {A.~V.}\
  \bibnamefont {Volotka}}, \bibinfo {author} {\bibfnamefont {C.~H.}\
  \bibnamefont {Keitel}}, \bibinfo {author} {\bibfnamefont {K.}~\bibnamefont
  {Blaum}}, \ and\ \bibinfo {author} {\bibfnamefont {S.}~\bibnamefont
  {Sturm}},\ }\href {\doibase 10.1038/s41586-022-04807-w} {\bibfield  {journal}
  {\bibinfo  {journal} {Nature}\ }\textbf {\bibinfo {volume} {606}},\ \bibinfo
  {pages} {479} (\bibinfo {year} {2022})}\BibitemShut {NoStop}%
\bibitem [{\citenamefont {Pachucki}\ \emph {et~al.}(2004)\citenamefont
  {Pachucki}, \citenamefont {Jentschura},\ and\ \citenamefont
  {Yerokhin}}]{Pachucki2004}%
  \BibitemOpen
  \bibfield  {author} {\bibinfo {author} {\bibfnamefont {K.}~\bibnamefont
  {Pachucki}}, \bibinfo {author} {\bibfnamefont {U.~D.}\ \bibnamefont
  {Jentschura}}, \ and\ \bibinfo {author} {\bibfnamefont {V.~A.}\ \bibnamefont
  {Yerokhin}},\ }\href {\doibase 10.1103/PhysRevLett.93.150401} {\bibfield
  {journal} {\bibinfo  {journal} {Phys. Rev. Lett.}\ }\textbf {\bibinfo
  {volume} {93}},\ \bibinfo {pages} {150401} (\bibinfo {year}
  {2004})}\BibitemShut {NoStop}%
\bibitem [{\citenamefont {Pachucki}\ and\ \citenamefont
  {Puchalski}(2017)}]{Pachucki2017}%
  \BibitemOpen
  \bibfield  {author} {\bibinfo {author} {\bibfnamefont {K.}~\bibnamefont
  {Pachucki}}\ and\ \bibinfo {author} {\bibfnamefont {M.}~\bibnamefont
  {Puchalski}},\ }\href {\doibase 10.1103/PhysRevA.96.032503} {\bibfield
  {journal} {\bibinfo  {journal} {Phys. Rev. A}\ }\textbf {\bibinfo {volume}
  {96}},\ \bibinfo {pages} {032503} (\bibinfo {year} {2017})}\BibitemShut
  {NoStop}%
\bibitem [{\citenamefont {Yerokhin}\ and\ \citenamefont
  {Harman}(2017)}]{Yerokhin2017}%
  \BibitemOpen
  \bibfield  {author} {\bibinfo {author} {\bibfnamefont {V.~A.}\ \bibnamefont
  {Yerokhin}}\ and\ \bibinfo {author} {\bibfnamefont {Z.}~\bibnamefont
  {Harman}},\ }\href {\doibase 10.1103/PhysRevA.95.060501} {\bibfield
  {journal} {\bibinfo  {journal} {Phys. Rev. A}\ }\textbf {\bibinfo {volume}
  {95}},\ \bibinfo {pages} {060501} (\bibinfo {year} {2017})}\BibitemShut
  {NoStop}%
\bibitem [{\citenamefont {Karshenboim}\ \emph {et~al.}(2005)\citenamefont
  {Karshenboim}, \citenamefont {Lee},\ and\ \citenamefont
  {Milstein}}]{Karshenboim2005}%
  \BibitemOpen
  \bibfield  {author} {\bibinfo {author} {\bibfnamefont {S.~G.}\ \bibnamefont
  {Karshenboim}}, \bibinfo {author} {\bibfnamefont {R.~N.}\ \bibnamefont
  {Lee}}, \ and\ \bibinfo {author} {\bibfnamefont {A.~I.}\ \bibnamefont
  {Milstein}},\ }\href {\doibase 10.1103/PhysRevA.72.042101} {\bibfield
  {journal} {\bibinfo  {journal} {Phys. Rev. A}\ }\textbf {\bibinfo {volume}
  {72}},\ \bibinfo {pages} {042101} (\bibinfo {year} {2005})}\BibitemShut
  {NoStop}%
\bibitem [{\citenamefont {Shabaev}(2002)}]{Shabaev2002}%
  \BibitemOpen
  \bibfield  {author} {\bibinfo {author} {\bibfnamefont {V.~M.}\ \bibnamefont
  {Shabaev}},\ }\href {\doibase https://doi.org/10.1016/S0370-1573(01)00024-2}
  {\bibfield  {journal} {\bibinfo  {journal} {Phys. Rep.}\ }\textbf {\bibinfo
  {volume} {356}},\ \bibinfo {pages} {119 } (\bibinfo {year}
  {2002})}\BibitemShut {NoStop}%
\bibitem [{\citenamefont {Beier}\ \emph {et~al.}(2000)\citenamefont {Beier},
  \citenamefont {Lindgren}, \citenamefont {Persson}, \citenamefont
  {Salomonson}, \citenamefont {Sunnergren}, \citenamefont {H\"affner},\ and\
  \citenamefont {Hermanspahn}}]{Beier2000}%
  \BibitemOpen
  \bibfield  {author} {\bibinfo {author} {\bibfnamefont {T.}~\bibnamefont
  {Beier}}, \bibinfo {author} {\bibfnamefont {I.}~\bibnamefont {Lindgren}},
  \bibinfo {author} {\bibfnamefont {H.}~\bibnamefont {Persson}}, \bibinfo
  {author} {\bibfnamefont {S.}~\bibnamefont {Salomonson}}, \bibinfo {author}
  {\bibfnamefont {P.}~\bibnamefont {Sunnergren}}, \bibinfo {author}
  {\bibfnamefont {H.}~\bibnamefont {H\"affner}}, \ and\ \bibinfo {author}
  {\bibfnamefont {N.}~\bibnamefont {Hermanspahn}},\ }\href {\doibase
  10.1103/PhysRevA.62.032510} {\bibfield  {journal} {\bibinfo  {journal} {Phys.
  Rev. A}\ }\textbf {\bibinfo {volume} {62}},\ \bibinfo {pages} {032510}
  (\bibinfo {year} {2000})}\BibitemShut {NoStop}%
\bibitem [{\citenamefont {Beier}(2000)}]{Beier2000-1}%
  \BibitemOpen
  \bibfield  {author} {\bibinfo {author} {\bibfnamefont {T.}~\bibnamefont
  {Beier}},\ }\href {\doibase https://doi.org/10.1016/S0370-1573(00)00071-5}
  {\bibfield  {journal} {\bibinfo  {journal} {Phys. Rep.}\ }\textbf {\bibinfo
  {volume} {339}},\ \bibinfo {pages} {79 } (\bibinfo {year}
  {2000})}\BibitemShut {NoStop}%
\bibitem [{\citenamefont {Czarnecki}\ \emph {et~al.}(2018)\citenamefont
  {Czarnecki}, \citenamefont {Dowling}, \citenamefont {Piclum},\ and\
  \citenamefont {Szafron}}]{Czarnecki2018}%
  \BibitemOpen
  \bibfield  {author} {\bibinfo {author} {\bibfnamefont {A.}~\bibnamefont
  {Czarnecki}}, \bibinfo {author} {\bibfnamefont {M.}~\bibnamefont {Dowling}},
  \bibinfo {author} {\bibfnamefont {J.}~\bibnamefont {Piclum}}, \ and\ \bibinfo
  {author} {\bibfnamefont {R.}~\bibnamefont {Szafron}},\ }\href {\doibase
  10.1103/PhysRevLett.120.043203} {\bibfield  {journal} {\bibinfo  {journal}
  {Phys. Rev. Lett.}\ }\textbf {\bibinfo {volume} {120}},\ \bibinfo {pages}
  {043203} (\bibinfo {year} {2018})}\BibitemShut {NoStop}%
\bibitem [{\citenamefont {Glazov}\ \emph {et~al.}(2019)\citenamefont {Glazov},
  \citenamefont {K\"ohler-Langes}, \citenamefont {Volotka}, \citenamefont
  {Blaum}, \citenamefont {Hei\ss{}e}, \citenamefont {Plunien}, \citenamefont
  {Quint}, \citenamefont {Rau}, \citenamefont {Shabaev}, \citenamefont
  {Sturm},\ and\ \citenamefont {Werth}}]{Glazov2019}%
  \BibitemOpen
  \bibfield  {author} {\bibinfo {author} {\bibfnamefont {D.~A.}\ \bibnamefont
  {Glazov}}, \bibinfo {author} {\bibfnamefont {F.}~\bibnamefont
  {K\"ohler-Langes}}, \bibinfo {author} {\bibfnamefont {A.~V.}\ \bibnamefont
  {Volotka}}, \bibinfo {author} {\bibfnamefont {K.}~\bibnamefont {Blaum}},
  \bibinfo {author} {\bibfnamefont {F.}~\bibnamefont {Hei\ss{}e}}, \bibinfo
  {author} {\bibfnamefont {G.}~\bibnamefont {Plunien}}, \bibinfo {author}
  {\bibfnamefont {W.}~\bibnamefont {Quint}}, \bibinfo {author} {\bibfnamefont
  {S.}~\bibnamefont {Rau}}, \bibinfo {author} {\bibfnamefont {V.~M.}\
  \bibnamefont {Shabaev}}, \bibinfo {author} {\bibfnamefont {S.}~\bibnamefont
  {Sturm}}, \ and\ \bibinfo {author} {\bibfnamefont {G.}~\bibnamefont
  {Werth}},\ }\href {\doibase 10.1103/PhysRevLett.123.173001} {\bibfield
  {journal} {\bibinfo  {journal} {Phys. Rev. Lett.}\ }\textbf {\bibinfo
  {volume} {123}},\ \bibinfo {pages} {173001} (\bibinfo {year}
  {2019})}\BibitemShut {NoStop}%
\bibitem [{\citenamefont {Sikora}\ \emph {et~al.}(2025)\citenamefont {Sikora},
  \citenamefont {Yerokhin}, \citenamefont {Keitel},\ and\ \citenamefont
  {Harman}}]{Sikora2025}%
  \BibitemOpen
  \bibfield  {author} {\bibinfo {author} {\bibfnamefont {B.}~\bibnamefont
  {Sikora}}, \bibinfo {author} {\bibfnamefont {V.~A.}\ \bibnamefont
  {Yerokhin}}, \bibinfo {author} {\bibfnamefont {C.~H.}\ \bibnamefont
  {Keitel}}, \ and\ \bibinfo {author} {\bibfnamefont {Z.}~\bibnamefont
  {Harman}},\ }\href {\doibase 10.1103/PhysRevLett.134.123001} {\bibfield
  {journal} {\bibinfo  {journal} {Phys. Rev. Lett.}\ }\textbf {\bibinfo
  {volume} {134}},\ \bibinfo {pages} {123001} (\bibinfo {year}
  {2025})}\BibitemShut {NoStop}%
\bibitem [{\citenamefont {Zatorski}\ \emph {et~al.}(2012)\citenamefont
  {Zatorski}, \citenamefont {Oreshkina}, \citenamefont {Keitel},\ and\
  \citenamefont {Harman}}]{Zatorski2012}%
  \BibitemOpen
  \bibfield  {author} {\bibinfo {author} {\bibfnamefont {J.}~\bibnamefont
  {Zatorski}}, \bibinfo {author} {\bibfnamefont {N.~S.}\ \bibnamefont
  {Oreshkina}}, \bibinfo {author} {\bibfnamefont {C.~H.}\ \bibnamefont
  {Keitel}}, \ and\ \bibinfo {author} {\bibfnamefont {Z.}~\bibnamefont
  {Harman}},\ }\href {\doibase 10.1103/PhysRevLett.108.063005} {\bibfield
  {journal} {\bibinfo  {journal} {Phys. Rev. Lett.}\ }\textbf {\bibinfo
  {volume} {108}},\ \bibinfo {pages} {063005} (\bibinfo {year}
  {2012})}\BibitemShut {NoStop}%
\bibitem [{\citenamefont {Srednicki}(2007)}]{Srednicki2007}%
  \BibitemOpen
  \bibfield  {author} {\bibinfo {author} {\bibfnamefont {M.}~\bibnamefont
  {Srednicki}},\ }\href@noop {} {\emph {\bibinfo {title} {Quantum Field
  Theory}}}\ (\bibinfo  {publisher} {Cambridge University Press},\ \bibinfo
  {year} {2007})\BibitemShut {NoStop}%
\bibitem [{\citenamefont {Debierre}\ \emph {et~al.}(2020)\citenamefont
  {Debierre}, \citenamefont {Keitel},\ and\ \citenamefont
  {Harman}}]{Debierre2020}%
  \BibitemOpen
  \bibfield  {author} {\bibinfo {author} {\bibfnamefont {V.}~\bibnamefont
  {Debierre}}, \bibinfo {author} {\bibfnamefont {C.~H.}\ \bibnamefont
  {Keitel}}, \ and\ \bibinfo {author} {\bibfnamefont {Z.}~\bibnamefont
  {Harman}},\ }\href {\doibase 10.1016/j.physletb.2020.135527} {\bibfield
  {journal} {\bibinfo  {journal} {Phys. Lett. B}\ }\textbf {\bibinfo {volume}
  {807}},\ \bibinfo {pages} {135527} (\bibinfo {year} {2020})}\BibitemShut
  {NoStop}%
\bibitem [{\citenamefont {Hegstrom}(1973)}]{Hegstrom1973}%
  \BibitemOpen
  \bibfield  {author} {\bibinfo {author} {\bibfnamefont {R.~A.}\ \bibnamefont
  {Hegstrom}},\ }\href {\doibase 10.1103/PhysRevA.7.451} {\bibfield  {journal}
  {\bibinfo  {journal} {Phys. Rev. A}\ }\textbf {\bibinfo {volume} {7}},\
  \bibinfo {pages} {451} (\bibinfo {year} {1973})}\BibitemShut {NoStop}%
\bibitem [{\citenamefont {Cakir}\ \emph {et~al.}(2020)\citenamefont {Cakir},
  \citenamefont {Yerokhin}, \citenamefont {Oreshkina}, \citenamefont {Sikora},
  \citenamefont {Tupitsyn}, \citenamefont {Keitel},\ and\ \citenamefont
  {Harman}}]{Cakir2020}%
  \BibitemOpen
  \bibfield  {author} {\bibinfo {author} {\bibfnamefont {H.}~\bibnamefont
  {Cakir}}, \bibinfo {author} {\bibfnamefont {V.~A.}\ \bibnamefont {Yerokhin}},
  \bibinfo {author} {\bibfnamefont {N.~S.}\ \bibnamefont {Oreshkina}}, \bibinfo
  {author} {\bibfnamefont {B.}~\bibnamefont {Sikora}}, \bibinfo {author}
  {\bibfnamefont {I.~I.}\ \bibnamefont {Tupitsyn}}, \bibinfo {author}
  {\bibfnamefont {C.~H.}\ \bibnamefont {Keitel}}, \ and\ \bibinfo {author}
  {\bibfnamefont {Z.}~\bibnamefont {Harman}},\ }\href {\doibase
  10.1103/PhysRevA.101.062513} {\bibfield  {journal} {\bibinfo  {journal}
  {Phys. Rev. A}\ }\textbf {\bibinfo {volume} {101}},\ \bibinfo {pages}
  {062513} (\bibinfo {year} {2020})}\BibitemShut {NoStop}%
\bibitem [{\citenamefont {Pachucki}\ \emph {et~al.}(2005)\citenamefont
  {Pachucki}, \citenamefont {Czarnecki}, \citenamefont {Jentschura},\ and\
  \citenamefont {Yerokhin}}]{Pachucki2005}%
  \BibitemOpen
  \bibfield  {author} {\bibinfo {author} {\bibfnamefont {K.}~\bibnamefont
  {Pachucki}}, \bibinfo {author} {\bibfnamefont {A.}~\bibnamefont {Czarnecki}},
  \bibinfo {author} {\bibfnamefont {U.~D.}\ \bibnamefont {Jentschura}}, \ and\
  \bibinfo {author} {\bibfnamefont {V.~A.}\ \bibnamefont {Yerokhin}},\ }\href
  {\doibase 10.1103/PhysRevA.72.022108} {\bibfield  {journal} {\bibinfo
  {journal} {Phys. Rev. A}\ }\textbf {\bibinfo {volume} {72}},\ \bibinfo
  {pages} {022108} (\bibinfo {year} {2005})}\BibitemShut {NoStop}%
\bibitem [{\citenamefont {Door}\ \emph {et~al.}(2025)\citenamefont {Door},
  \citenamefont {Yeh}, \citenamefont {Heinz}, \citenamefont {Kirk},
  \citenamefont {Lyu}, \citenamefont {Miyagi}, \citenamefont {Berengut},
  \citenamefont {Biero\ifmmode~\acute{n}\else \'{n}\fi{}}, \citenamefont
  {Blaum}, \citenamefont {Dreissen}, \citenamefont {Eliseev}, \citenamefont
  {Filianin}, \citenamefont {Filzinger}, \citenamefont {Fuchs}, \citenamefont
  {F\"urst}, \citenamefont {Gaigalas}, \citenamefont {Harman}, \citenamefont
  {Herkenhoff}, \citenamefont {Huntemann}, \citenamefont {Keitel},
  \citenamefont {Kromer}, \citenamefont {Lange}, \citenamefont {Rischka},
  \citenamefont {Schweiger}, \citenamefont {Schwenk}, \citenamefont {Shimizu},\
  and\ \citenamefont {Mehlst\"aubler}}]{Door2025}%
  \BibitemOpen
  \bibfield  {author} {\bibinfo {author} {\bibfnamefont {M.}~\bibnamefont
  {Door}}, \bibinfo {author} {\bibfnamefont {C.-H.}\ \bibnamefont {Yeh}},
  \bibinfo {author} {\bibfnamefont {M.}~\bibnamefont {Heinz}}, \bibinfo
  {author} {\bibfnamefont {F.}~\bibnamefont {Kirk}}, \bibinfo {author}
  {\bibfnamefont {C.}~\bibnamefont {Lyu}}, \bibinfo {author} {\bibfnamefont
  {T.}~\bibnamefont {Miyagi}}, \bibinfo {author} {\bibfnamefont {J.~C.}\
  \bibnamefont {Berengut}}, \bibinfo {author} {\bibfnamefont {J.}~\bibnamefont
  {Biero\ifmmode~\acute{n}\else \'{n}\fi{}}}, \bibinfo {author} {\bibfnamefont
  {K.}~\bibnamefont {Blaum}}, \bibinfo {author} {\bibfnamefont {L.~S.}\
  \bibnamefont {Dreissen}}, \bibinfo {author} {\bibfnamefont {S.}~\bibnamefont
  {Eliseev}}, \bibinfo {author} {\bibfnamefont {P.}~\bibnamefont {Filianin}},
  \bibinfo {author} {\bibfnamefont {M.}~\bibnamefont {Filzinger}}, \bibinfo
  {author} {\bibfnamefont {E.}~\bibnamefont {Fuchs}}, \bibinfo {author}
  {\bibfnamefont {H.~A.}\ \bibnamefont {F\"urst}}, \bibinfo {author}
  {\bibfnamefont {G.}~\bibnamefont {Gaigalas}}, \bibinfo {author}
  {\bibfnamefont {Z.}~\bibnamefont {Harman}}, \bibinfo {author} {\bibfnamefont
  {J.}~\bibnamefont {Herkenhoff}}, \bibinfo {author} {\bibfnamefont
  {N.}~\bibnamefont {Huntemann}}, \bibinfo {author} {\bibfnamefont {C.~H.}\
  \bibnamefont {Keitel}}, \bibinfo {author} {\bibfnamefont {K.}~\bibnamefont
  {Kromer}}, \bibinfo {author} {\bibfnamefont {D.}~\bibnamefont {Lange}},
  \bibinfo {author} {\bibfnamefont {A.}~\bibnamefont {Rischka}}, \bibinfo
  {author} {\bibfnamefont {C.}~\bibnamefont {Schweiger}}, \bibinfo {author}
  {\bibfnamefont {A.}~\bibnamefont {Schwenk}}, \bibinfo {author} {\bibfnamefont
  {N.}~\bibnamefont {Shimizu}}, \ and\ \bibinfo {author} {\bibfnamefont
  {T.~E.}\ \bibnamefont {Mehlst\"aubler}},\ }\href {\doibase
  10.1103/PhysRevLett.134.063002} {\bibfield  {journal} {\bibinfo  {journal}
  {Phys. Rev. Lett.}\ }\textbf {\bibinfo {volume} {134}},\ \bibinfo {pages}
  {063002} (\bibinfo {year} {2025})}\BibitemShut {NoStop}%
\bibitem [{\citenamefont {Ono}\ \emph {et~al.}(2022)\citenamefont {Ono},
  \citenamefont {Saito}, \citenamefont {Ishiyama}, \citenamefont {Higomoto},
  \citenamefont {Takano}, \citenamefont {Takasu}, \citenamefont {Yamamoto},
  \citenamefont {Tanaka},\ and\ \citenamefont {Takahashi}}]{Ono2022}%
  \BibitemOpen
  \bibfield  {author} {\bibinfo {author} {\bibfnamefont {K.}~\bibnamefont
  {Ono}}, \bibinfo {author} {\bibfnamefont {Y.}~\bibnamefont {Saito}}, \bibinfo
  {author} {\bibfnamefont {T.}~\bibnamefont {Ishiyama}}, \bibinfo {author}
  {\bibfnamefont {T.}~\bibnamefont {Higomoto}}, \bibinfo {author}
  {\bibfnamefont {T.}~\bibnamefont {Takano}}, \bibinfo {author} {\bibfnamefont
  {Y.}~\bibnamefont {Takasu}}, \bibinfo {author} {\bibfnamefont
  {Y.}~\bibnamefont {Yamamoto}}, \bibinfo {author} {\bibfnamefont
  {M.}~\bibnamefont {Tanaka}}, \ and\ \bibinfo {author} {\bibfnamefont
  {Y.}~\bibnamefont {Takahashi}},\ }\href {\doibase 10.1103/PhysRevX.12.021033}
  {\bibfield  {journal} {\bibinfo  {journal} {Phys. Rev. X}\ }\textbf {\bibinfo
  {volume} {12}},\ \bibinfo {pages} {021033} (\bibinfo {year}
  {2022})}\BibitemShut {NoStop}%
\bibitem [{\citenamefont {Solaro}\ \emph {et~al.}(2020)\citenamefont {Solaro},
  \citenamefont {Meyer}, \citenamefont {Fisher}, \citenamefont {Berengut},
  \citenamefont {Fuchs},\ and\ \citenamefont {Drewsen}}]{Solaro2020}%
  \BibitemOpen
  \bibfield  {author} {\bibinfo {author} {\bibfnamefont {C.}~\bibnamefont
  {Solaro}}, \bibinfo {author} {\bibfnamefont {S.}~\bibnamefont {Meyer}},
  \bibinfo {author} {\bibfnamefont {K.}~\bibnamefont {Fisher}}, \bibinfo
  {author} {\bibfnamefont {J.~C.}\ \bibnamefont {Berengut}}, \bibinfo {author}
  {\bibfnamefont {E.}~\bibnamefont {Fuchs}}, \ and\ \bibinfo {author}
  {\bibfnamefont {M.}~\bibnamefont {Drewsen}},\ }\href {\doibase
  10.1103/PhysRevLett.125.123003} {\bibfield  {journal} {\bibinfo  {journal}
  {Phys. Rev. Lett.}\ }\textbf {\bibinfo {volume} {125}},\ \bibinfo {pages}
  {123003} (\bibinfo {year} {2020})}\BibitemShut {NoStop}%
\bibitem [{Sup()}]{Supplemental_Material}%
  \BibitemOpen
  \href@noop {} {}\bibinfo {note} {See Supplemental Material for details about
  the construction of blue and green dashed lines of Fig.~2}\BibitemShut
  {NoStop}%
\bibitem [{\citenamefont {Yerokhin}\ \emph {et~al.}(2012)\citenamefont
  {Yerokhin}, \citenamefont {Pachucki}, \citenamefont {Harman},\ and\
  \citenamefont {Keitel}}]{Yerokhin2012}%
  \BibitemOpen
  \bibfield  {author} {\bibinfo {author} {\bibfnamefont {V.~A.}\ \bibnamefont
  {Yerokhin}}, \bibinfo {author} {\bibfnamefont {K.}~\bibnamefont {Pachucki}},
  \bibinfo {author} {\bibfnamefont {Z.}~\bibnamefont {Harman}}, \ and\ \bibinfo
  {author} {\bibfnamefont {C.~H.}\ \bibnamefont {Keitel}},\ }\href {\doibase
  10.1103/PhysRevA.85.022512} {\bibfield  {journal} {\bibinfo  {journal} {Phys.
  Rev. A}\ }\textbf {\bibinfo {volume} {85}},\ \bibinfo {pages} {022512}
  (\bibinfo {year} {2012})}\BibitemShut {NoStop}%
\bibitem [{\citenamefont {Moskovkin}\ and\ \citenamefont
  {Shabaev}(2006)}]{Moskovkin2006}%
  \BibitemOpen
  \bibfield  {author} {\bibinfo {author} {\bibfnamefont {D.~L.}\ \bibnamefont
  {Moskovkin}}\ and\ \bibinfo {author} {\bibfnamefont {V.~M.}\ \bibnamefont
  {Shabaev}},\ }\href {\doibase 10.1103/PhysRevA.73.052506} {\bibfield
  {journal} {\bibinfo  {journal} {Phys. Rev. A}\ }\textbf {\bibinfo {volume}
  {73}},\ \bibinfo {pages} {052506} (\bibinfo {year} {2006})}\BibitemShut
  {NoStop}%
\bibitem [{\citenamefont {Sikora}(2023)}]{Sikora_private_comm2023}%
  \BibitemOpen
  \bibfield  {author} {\bibinfo {author} {\bibfnamefont {B.}~\bibnamefont
  {Sikora}},\ }\href@noop {} {\enquote {\bibinfo {title} {Private
  communication},}\ } (\bibinfo {year} {2023})\BibitemShut {NoStop}%
\bibitem [{\citenamefont {Angeli}\ and\ \citenamefont
  {Marinova}(2013)}]{Angeli2013}%
  \BibitemOpen
  \bibfield  {author} {\bibinfo {author} {\bibfnamefont {I.}~\bibnamefont
  {Angeli}}\ and\ \bibinfo {author} {\bibfnamefont {K.}~\bibnamefont
  {Marinova}},\ }\href {\doibase https://doi.org/10.1016/j.adt.2011.12.006}
  {\bibfield  {journal} {\bibinfo  {journal} {At. Data Nucl. Data Tables}\
  }\textbf {\bibinfo {volume} {99}},\ \bibinfo {pages} {69 } (\bibinfo {year}
  {2013})}\BibitemShut {NoStop}%
\bibitem [{\citenamefont {Matveev}\ \emph {et~al.}(2013)\citenamefont
  {Matveev}, \citenamefont {Parthey}, \citenamefont {Predehl}, \citenamefont
  {Alnis}, \citenamefont {Beyer}, \citenamefont {Holzwarth}, \citenamefont
  {Udem}, \citenamefont {Wilken}, \citenamefont {Kolachevsky}, \citenamefont
  {Abgrall}, \citenamefont {Rovera}, \citenamefont {Salomon}, \citenamefont
  {Laurent}, \citenamefont {Grosche}, \citenamefont {Terra}, \citenamefont
  {Legero}, \citenamefont {Schnatz}, \citenamefont {Weyers}, \citenamefont
  {Altschul},\ and\ \citenamefont {H\"ansch}}]{Matveev2013}%
  \BibitemOpen
  \bibfield  {author} {\bibinfo {author} {\bibfnamefont {A.}~\bibnamefont
  {Matveev}}, \bibinfo {author} {\bibfnamefont {C.~G.}\ \bibnamefont
  {Parthey}}, \bibinfo {author} {\bibfnamefont {K.}~\bibnamefont {Predehl}},
  \bibinfo {author} {\bibfnamefont {J.}~\bibnamefont {Alnis}}, \bibinfo
  {author} {\bibfnamefont {A.}~\bibnamefont {Beyer}}, \bibinfo {author}
  {\bibfnamefont {R.}~\bibnamefont {Holzwarth}}, \bibinfo {author}
  {\bibfnamefont {T.}~\bibnamefont {Udem}}, \bibinfo {author} {\bibfnamefont
  {T.}~\bibnamefont {Wilken}}, \bibinfo {author} {\bibfnamefont
  {N.}~\bibnamefont {Kolachevsky}}, \bibinfo {author} {\bibfnamefont
  {M.}~\bibnamefont {Abgrall}}, \bibinfo {author} {\bibfnamefont
  {D.}~\bibnamefont {Rovera}}, \bibinfo {author} {\bibfnamefont
  {C.}~\bibnamefont {Salomon}}, \bibinfo {author} {\bibfnamefont
  {P.}~\bibnamefont {Laurent}}, \bibinfo {author} {\bibfnamefont
  {G.}~\bibnamefont {Grosche}}, \bibinfo {author} {\bibfnamefont
  {O.}~\bibnamefont {Terra}}, \bibinfo {author} {\bibfnamefont
  {T.}~\bibnamefont {Legero}}, \bibinfo {author} {\bibfnamefont
  {H.}~\bibnamefont {Schnatz}}, \bibinfo {author} {\bibfnamefont
  {S.}~\bibnamefont {Weyers}}, \bibinfo {author} {\bibfnamefont
  {B.}~\bibnamefont {Altschul}}, \ and\ \bibinfo {author} {\bibfnamefont
  {T.~W.}\ \bibnamefont {H\"ansch}},\ }\href {\doibase
  10.1103/PhysRevLett.110.230801} {\bibfield  {journal} {\bibinfo  {journal}
  {Phys. Rev. Lett.}\ }\textbf {\bibinfo {volume} {110}},\ \bibinfo {pages}
  {230801} (\bibinfo {year} {2013})}\BibitemShut {NoStop}%
\bibitem [{\citenamefont {Yerokhin}\ and\ \citenamefont
  {Shabaev}(2015)}]{Yerokhin2015}%
  \BibitemOpen
  \bibfield  {author} {\bibinfo {author} {\bibfnamefont {V.~A.}\ \bibnamefont
  {Yerokhin}}\ and\ \bibinfo {author} {\bibfnamefont {V.~M.}\ \bibnamefont
  {Shabaev}},\ }\href {\doibase 10.1063/1.4927487} {\bibfield  {journal}
  {\bibinfo  {journal} {Journal of Physical and Chemical Reference Data}\
  }\textbf {\bibinfo {volume} {44}},\ \bibinfo {pages} {033103} (\bibinfo
  {year} {2015})},\ \Eprint
  {http://arxiv.org/abs/https://pubs.aip.org/aip/jpr/article-pdf/doi/10.1063/1.4927487/15981145/033103\_1\_online.pdf}
  {https://pubs.aip.org/aip/jpr/article-pdf/doi/10.1063/1.4927487/15981145/033103\_1\_online.pdf}
  \BibitemShut {NoStop}%
\bibitem [{\citenamefont {Potvliege}(2025)}]{Potvliege2025}%
  \BibitemOpen
  \bibfield  {author} {\bibinfo {author} {\bibfnamefont {R.~M.}\ \bibnamefont
  {Potvliege}},\ }\href {\doibase 10.1088/1367-2630/adc625} {\bibfield
  {journal} {\bibinfo  {journal} {New Journal of Physics}\ }\textbf {\bibinfo
  {volume} {27}},\ \bibinfo {pages} {045002} (\bibinfo {year}
  {2025})}\BibitemShut {NoStop}%
\bibitem [{\citenamefont {Jones}\ \emph {et~al.}(2020)\citenamefont {Jones},
  \citenamefont {Potvliege},\ and\ \citenamefont {Spannowsky}}]{Jones2020}%
  \BibitemOpen
  \bibfield  {author} {\bibinfo {author} {\bibfnamefont {M.~P.~A.}\
  \bibnamefont {Jones}}, \bibinfo {author} {\bibfnamefont {R.~M.}\ \bibnamefont
  {Potvliege}}, \ and\ \bibinfo {author} {\bibfnamefont {M.}~\bibnamefont
  {Spannowsky}},\ }\href {\doibase 10.1103/PhysRevResearch.2.013244} {\bibfield
   {journal} {\bibinfo  {journal} {Phys. Rev. Res.}\ }\textbf {\bibinfo
  {volume} {2}},\ \bibinfo {pages} {013244} (\bibinfo {year}
  {2020})}\BibitemShut {NoStop}%
\bibitem [{\citenamefont {Morgner}\ \emph {et~al.}(2023)\citenamefont
  {Morgner}, \citenamefont {Tu}, \citenamefont {M.}, \citenamefont {Sailer},
  \citenamefont {Heisse}, \citenamefont {Bekker}, \citenamefont {Sikora},
  \citenamefont {Lyu}, \citenamefont {Yerokhin}, \citenamefont {Harman},
  \citenamefont {Crespo López-Urrutia}, \citenamefont {Keitel}, \citenamefont
  {Sturm},\ and\ \citenamefont {Blaum}}]{Morgner2023}%
  \BibitemOpen
  \bibfield  {author} {\bibinfo {author} {\bibfnamefont {J.}~\bibnamefont
  {Morgner}}, \bibinfo {author} {\bibfnamefont {B.}~\bibnamefont {Tu}},
  \bibinfo {author} {\bibfnamefont {K.~C.}\ \bibnamefont {M.}}, \bibinfo
  {author} {\bibfnamefont {T.}~\bibnamefont {Sailer}}, \bibinfo {author}
  {\bibfnamefont {F.}~\bibnamefont {Heisse}}, \bibinfo {author} {\bibfnamefont
  {H.}~\bibnamefont {Bekker}}, \bibinfo {author} {\bibfnamefont
  {B.}~\bibnamefont {Sikora}}, \bibinfo {author} {\bibfnamefont
  {C.}~\bibnamefont {Lyu}}, \bibinfo {author} {\bibfnamefont {V.~A.}\
  \bibnamefont {Yerokhin}}, \bibinfo {author} {\bibfnamefont {Z.}~\bibnamefont
  {Harman}}, \bibinfo {author} {\bibfnamefont {J.~R.}\ \bibnamefont {Crespo
  López-Urrutia}}, \bibinfo {author} {\bibfnamefont {C.~H.}\ \bibnamefont
  {Keitel}}, \bibinfo {author} {\bibfnamefont {S.}~\bibnamefont {Sturm}}, \
  and\ \bibinfo {author} {\bibfnamefont {K.}~\bibnamefont {Blaum}},\ }\href
  {\doibase 10.1038/s41586-023-06453-2} {\bibfield  {journal} {\bibinfo
  {journal} {Nature}\ }\textbf {\bibinfo {volume} {622}},\ \bibinfo {pages}
  {53} (\bibinfo {year} {2023})}\BibitemShut {NoStop}%
\bibitem [{\citenamefont {Sturm}(2025)}]{Sturm_private_comm2025}%
  \BibitemOpen
  \bibfield  {author} {\bibinfo {author} {\bibfnamefont {S.}~\bibnamefont
  {Sturm}},\ }\href@noop {} {\enquote {\bibinfo {title} {Private
  communication},}\ } (\bibinfo {year} {2025})\BibitemShut {NoStop}%
\end{thebibliography}
\end{document}


\title{\Large Supplemental material: Fermion-selective tests of new physics with the bound-electron $g$ factor}
\date{\today}
\author{M. Moretti}
\affiliation{Max-Planck-Institut für Kernphysik, Heidelberg, Germany}
\author{C. H. Keitel}
\affiliation{Max-Planck-Institut für Kernphysik, Heidelberg, Germany}
\author{Z. Harman}
\affiliation{Max-Planck-Institut für Kernphysik, Heidelberg, Germany}
\maketitle

\noindent Throughout this Supplement, we present mathematical details of the method put forward in the main body of the article, \emph{i.e.} we will explain how the blue and green dashed exclusion curves of Fig.~2 of the paper have been derived. In particular, we summarize how to isolate the contribution to the $g$ factor from the electron--proton interaction of an H-like ion.\\
We start by re-writing some useful formulae.
\begin{itemize}
    \item Bound on the contribution of the $g$ factor due to a new scalar particle:
    \begin{equation}
        \label{eq:gfactor bound}
        \Delta g_\phi \leq \Delta g^\text{exp} - \Delta g^\text{th} \, .
    \end{equation}
    Both theoretical and experimental results have error bars. When our scalar particle contributes negatively (positively), the allowed range for new physics is defined by the gap between the upper (lower) limit of the theoretical uncertainty and the lower (upper) limit of the experimental one.
    \item Total $g$ factor scalar correction:
    \begin{equation}
        \label{eq:total gfactor}
        \Delta g_\phi = \Delta g_\phi^{ee} + \Delta g_\phi^{en} + \Delta g_\phi^{ep}.
    \end{equation}
    Our model considers the general case in which the scalar interacts with all the fermions in the atom. Then the $g$ factor of the electron receives corrections from electron--electron (in a H-like ion, this is a self interaction), electron--neutron and electron--proton interactions.
    \item Leading order binding approximation to the electron--electron $g$ factor scalar correction:
    \begin{equation}
        \label{eq:g-factor SE first approx}
        \Delta g_\phi^{ee} = \Delta g_\phi^{ee,\text{free}} \left( 1 - \frac{(Z\alpha)^2}{6} \right) \, ,
    \end{equation}
    where the free part is $\Delta g_\phi^{ee,\text{free}} = 2 F_2^\phi (0)$, for which we obtain
    \begin{multline}
    \label{eq:F2(0)}
        F_2^\phi (0) = \frac{\alpha_{ee}}{2\pi} \left\{ \frac{3}{2} - r_m^2 -\frac{r_m \left( r_m^4 - 5r_m^2 +4 \right)}{\sqrt{r_m^2 - 4}} \, \times \right. \\
        \left. \times \left[ \text{atanh} \left( \frac{r_m^2 - 2}{r_m\sqrt{r_m^2 - 4}} \right) - \text{atanh} \left( \frac{r_m}{\sqrt{r_m^2 - 4}} \right) \right] + r_m^2 \left( r_m^2 - 3 \right) \text{ln}\left( r_m \right) \right\} \, ,
    \end{multline}
    where $r_m = m_\phi/m_e$ and $\alpha_{ee}=y_e^2/4\pi$.
    \item Electron--nucleon correction for 1$s$ electrons. Here, the capital $N$ indicates nucleons in general, and stands for the number of neutrons $A-Z$ or protons $Z$:
    \begin{align}
        \nonumber
        \Delta g_\phi^{eN} & = -\frac{4}{3}\alpha_{ep} \, N \, \frac{Z\alpha}{\gamma} \left( 1+ \frac{m_\phi}{2Z\alpha \, m_e} \right)^{-2\gamma} \left[ 3 - 2\frac{\left( Z\alpha \right)^2}{1+\gamma} - 2\gamma \left( 1+ \frac{m_\phi}{2Z\alpha \, m_e}  \right)^{-1}  \right] \\
        \label{eq:gfactor eN}
        & = \alpha_{eN} \, N \, f(Z) \, ,
    \end{align}
    where the $Z$-dependence has been completely enclosed into the function $f(Z)$.
\end{itemize}
Due to Eq. \eqref{eq:total gfactor}, the bosonic corrections to the $g$ factors of two different ions are given by
\begin{align}
\label{eq:g-factor ion1}
    & \Delta g_{\phi,1} = \Delta g_\phi^{ee,\text{free}} \left( 1 - \frac{(Z_1\alpha)^2}{6} \right) + \alpha_{en}(A_1-Z_1)f(Z_1) + \alpha_{ep} \, Z_1 \,f(Z_1) \, ,\\
\label{eq:g-factor ion2}
    & \Delta g_{\phi,2} = \Delta g_\phi^{ee,\text{free}} \left( 1 - \frac{(Z_2\alpha)^2}{6} \right) + \alpha_{en}(A_2-Z_2)f(Z_2) + \alpha_{ep} \, Z_2 \,f(Z_2) \, .
\end{align}
We now introduce $x=\frac{A_1-Z_1}{A_2-Z_2}\frac{f(Z_1)}{f(Z_2)}$ and $y=Z_1-Z_2\frac{A_1-Z_1}{A_2-Z_2}$, and define the following weighted difference: 
\begin{equation}
    \label{eq:weighted difference}
    \Delta g_{\phi,1} - x \, \Delta g_{\phi,2} = \Delta g_\phi^{ee,free} \left[ \left( 1-\frac{\left(Z_1\alpha\right)^2}{6} \right) - x \left( 1-\frac{\left(Z_2\alpha\right)^2}{6} \right) \right] + \alpha_{ep} \, y \, f(Z_1) \, ,
\end{equation}
where higher orders in $Z\alpha$ have been omitted.\\
This allows for a complete cancellation of the neutron contribution but also, assuming $Z_1$ close to $Z_2$, as long as $x \simeq 1$, an electron contribution suppression (We can see \emph{a posteriori} that the suppression is valid even if $x$ is not close to $1$ because the electron contribution bound is more competitive than the protonic one). As a consequence, we can focus on the proton contribution only.\\
To be able to draw exclusion lines, we need both theoretical and experimental data, each of which consists of the $g$ factor value and its uncertainty. Some reasonable assumptions can be made for uncertainties. In what follows, we assume that experimental and theoretical values are equal; in this manner, only error bars of the hypothetical data matter. Before explaining how we estimated the corresponding uncertainties, we want to clarify our choice of the element pairs used in Fig.~2 of the Letter.\\

\begin{figure}
    \centering
        \includegraphics[width=0.45\textwidth]{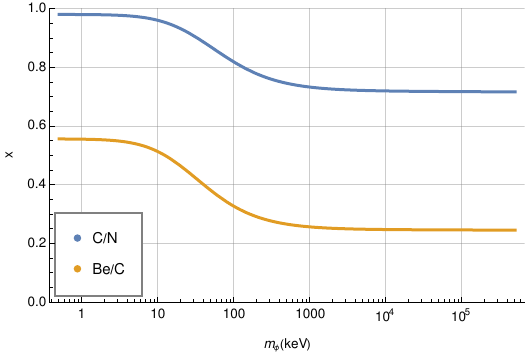}
    \caption{\justifying Two values for $x$, as a function of the scalar mass $m_\phi$ are shown. Blue and yellow lines represent the $^{14}\text{C}^{5+}/^{14}\text{N}^{6+}$ and $^{9}\text{Be}^{3+}/^{12}\text{C}^{5+}$ pairs, with values spanning in the ranges $[0.72,0.98]$ and $[0.25,0.56]$, respectively.}
    \label{fig:xPlots}
\end{figure}

\begin{figure}
    \centering
        \includegraphics[width=0.45\textwidth]{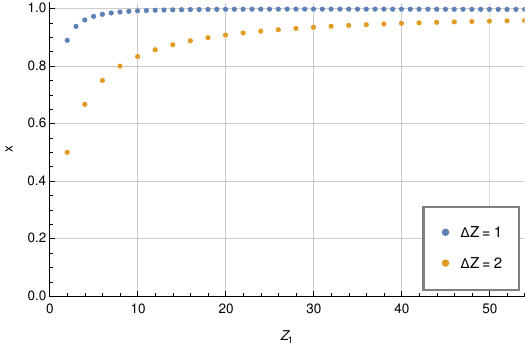}
    \caption{\justifying Two lists of values of $x$, for $\Delta A_1 = 2$ and $\Delta A_2 = 0$, are shown. For $\Delta Z = 2$ we draw only isotopes with even number of protons, because most of the times their nuclei are spinless, making measurements of $g$ factors easier. When instead one considers $\Delta Z = 1$, one of the two isotopes carries a nuclear spin; as a consequence, the $g$ factor of the electron receives some corrections depending on the nuclear $g$ factor which are well understood.}
    \label{fig:DeltaZ's}
\end{figure}

\paragraph{Choice of the element pairs}\,\\
The selection of the pair of ions for the nuclide shift is very broad, ideally speaking the number of pairs is the square of the number of isotopes of the whole periodic table. However, there are some guidelines that motivate our choices. First, we aim for not too heavy isotopes, because theoretical uncertainties grow quite fast with the atomic number; second, we intend to keep $x=\frac{A_1-Z_1}{A_2-Z_2}\frac{f(Z_1)}{f(Z_2)}$ as close as possible to $1$, in order to reduce the uncertainty of the experimental value of the weighted difference (see the next Section); finally, since a measurement may take weeks, we require quite some stability from our isotopes. Being $x$ a function of $m_\phi$, its value varies with the scalar mass. In Fig.~\ref{fig:xPlots} we draw $x$ for the pairs of ions used in the Letter. From the expression of $x$ we can see that there are four free parameters (technically speaking, once $Z_1$ and $Z_2$ are chosen, $A_1$ and $A_2$ are not completely free), and we take care of them in pairs. It may be useful to express the mass numbers as their difference with twice the atomic numbers $A_{1(2)} = 2Z_{1(2)} + \Delta A_{1(2)}$, then we recast $x$ by replacing $A_{1(2)}$ with its expression involving $\Delta A_{1(2)}$
\begin{equation}
    \label{eq:x}
    x = \frac{Z_1 + \Delta A_1}{Z_2 + \Delta A_2}\frac{f(Z_1)}{f(Z_2)}
\end{equation}
With this choice, we know that, with very few exceptions, stable isotopes carry $\Delta A \geq 0$.
\begin{itemize}
    \item Nuclear numbers $Z_1$, $Z_2$: the demand to keep $x\simeq 1$ fixes the atomic number difference $\Delta Z = Z_1 - Z_2$ to be either 1 or 2, otherwise $x$ decreases quickly as the atomic number approaches the origin (see Fig.~\ref{fig:DeltaZ's}).
    \item Atomic masses $\Delta A_1$, $\Delta A_2$: from expression \eqref{eq:x}, it can be seen that $x$ increases as $\Delta A_1$ increases and $\Delta A_2$ decreases. Advantageous values for these parameters, given our aim to focus on small atomic numbers, are $\Delta A_1 = 2$ and $\Delta A_2 = 0$.
\end{itemize}
Now we deal with the estimation of experimental and theoretical uncertainties. Even if this is not a strict requirement, we would like to have these to be comparable to each other.\\
Notice, in fact, that both contributions of the right hand side of Eq. \eqref{eq:gfactor bound} take the form of the left hand side of Eq. \eqref{eq:weighted difference}, for the nuclide shift, \emph{i.e.}:
\begin{align}
    \label{eq:weighted difference exp}
    & \Delta g^\text{exp} = \Delta g^\text{exp}_{\phi,1} - x \, \Delta g^\text{exp}_{\phi,2} \, , \\
    \label{eq:weighted difference th}
    & \Delta g^\text{th} = \Delta g^\text{th}_{\phi,1} - x \, \Delta g^\text{th}_{\phi,2} \, .
\end{align}
The difference between these latter two defines the allowed window of a possible scalar contribution.

\paragraph{Estimation of the experimental error bar}\,\\
Eq. \eqref{eq:weighted difference exp} can be suitably recast into
\begin{equation}
    \label{eq:weighted difference exp recast}
    \Delta g^\text{exp}_{\phi,1} - x \, \Delta g^\text{exp}_{\phi,2} = \Delta g^\text{exp}_{\phi,1,2} + (1-x) \Delta g^\text{exp}_{\phi,2} \, ,
\end{equation}
where we added and subtracted $\Delta g^\text{exp}_{\phi,2}$ and defined the simple difference $\Delta g^\text{exp}_{\phi,1,2} = \Delta g^\text{exp}_{\phi,1} -\Delta g^\text{exp}_{\phi,2}$. Given Eq. \eqref{eq:weighted difference exp recast}, we need two experimental quantities: the shift $\Delta g^\text{exp}_{\phi,1,2}$ and the $g$ factor of the second ion, $\Delta g^\text{exp}_{\phi,2}$. Since nuclide shift $g$ factor experiments have not yet been performed, by similarity, we assume its uncertainty to be that of the already measured isotope shift, $g \left( ^{20}\text{Ne}^{9+} \right) - g \left( ^{22}\text{Ne}^{9+} \right)$ in Ref. \cite{Sailer2022}, and for the second ion from the $g$ factor experiment of $^{20}\text{Ne}^{9+}$ in Ref. \cite{Heisse2023}. These uncertainties are of the order of $10^{-12}$ and $10^{-10}$, respectively. Then, if $x\simeq 0.99$, \emph{i.e.} $1-x \simeq 0.01$, the uncertainties of the terms in Eq. \eqref{eq:weighted difference exp recast} share the same order of magnitude of $10^{-12}$. \\
The role of $x$ is therefore to keep the uncertainty of the experimentally determined weighted difference sufficiently low.

\paragraph{Estimation of the theoretical uncertainty}\,\\
To estimate the theoretical uncertainty we rely on \cite{Czarnecki2018}, where the largest part of the error bar is accounted for as 
\begin{equation}
    \label{eq:Czarnecki error}
    \sigma = \left( \frac{\alpha}{\pi} \right)^2 (Z\alpha)^6\text{ln}^3\left((Z\alpha )^{-2}\right),
\end{equation}
as long as one considers ions whose atomic number is $Z \gtrsim 6$. Given this latter formula, it is clear that the uncertainties of the $g$ factors of a pair of ions, let they be $\sigma_1$ and $\sigma_2$, are related (through the dependence on the coefficient $Z\alpha$), then when computing the difference of the $g$ factors $\Delta g^{th} = g_1^{th} - x \, g_2^{th}$, the uncertainty of the result $\sigma_r$, instead of being computed as a plain or a quadratic sum, can be reduced to
\begin{equation}
    \label{eq:uncertainty propagated}
    \sigma_{r} = \sqrt{\sigma_1^2 + x^2\sigma_2^2 - 2x\sigma_{1,2}} \quad \text{, with} \quad \sigma_{1,2} = \frac{\partial\sigma_1}{\partial Z} \frac{\partial\sigma_2}{\partial Z} \sigma_Z^2 \quad \text{and} \quad \sigma_Z = 1.
\end{equation}
We used this relation to draw the green dashed line of Fig.~2 of the Letter. Instead, to draw the blue dashed one, we used a simple sum in quadrature of uncertainties, because data for beryllium and carbon stem from different sources~\cite{Sikora_private_comm2023,Czarnecki2018}, hence they come from different assumptions.
In both cases, the theoretical error bar was not below $10^{-11}$.\\
The black dashed line has been drawn in a way similar to the green one, but with the realistic assumption of an improvement of a factor of 10 in the experimental accuracy~\cite{Sturm_private_comm2025} with respect to the neon isotope shift~\cite{Sailer2022} and a matching improvement of the theory.
\\

In summary, the ultimate goal is to achieve uncertainties of the same order of magnitude for theoretical and experimental data. Having $x \simeq 1$ is useful in decreasing the experimental uncertainty of the $g$ factor of one of the nuclides [see Eq.~(\ref{eq:weighted difference exp recast})]. Nonetheless, with other values of $x$, still a competitive bound can be reached: e.g. the blue curve of Fig.~2 in the Letter, values $x \in [0.24,0.56]$ were used.